\documentclass[aps,prl,superscriptaddress,  reprint,nofootinbib]{revtex4-1}
 \usepackage{graphicx}
\usepackage{dcolumn}
\usepackage{bm}
\usepackage{amssymb,amsmath}
\usepackage{epstopdf}
\usepackage{color}
\usepackage{subfigure}

\begin{document}

\title{Modelling electron-phonon interactions in graphene with curved space hydrodynamics}

\author{I. Giordanelli} 
\email{gilario@ethz.ch} 
\affiliation{ ETH Z\"urich, Computational Physics for Engineering Materials, Institute for Building Materials, Wolfgang-Pauli-Strasse 27, HIT, CH-8093 Z\"urich (Switzerland)}
\author{M. Mendoza} 
\email{mmendoza@ethz.ch} 
\affiliation{ ETH Z\"urich, Computational Physics for Engineering Materials, Institute for Building Materials, Wolfgang-Pauli-Strasse 27, HIT, CH-8093 Z\"urich (Switzerland)}
\author{H. J. Herrmann}
\email{hjherrmann@ethz.ch} 
\affiliation{ ETH Z\"urich, Computational Physics for Engineering Materials, Institute for Building Materials, Wolfgang-Pauli-Strasse 27, HIT, CH-8093 Z\"urich (Switzerland)}
 \affiliation{Departamento de F\'isica, Universidade
   Federal do Cear\'a, Campus do Pici, 60455-760 Fortaleza, Cear\'a,
   (Brazil)}

\date{\today}

\begin{abstract}
We introduce a different perspective describing electron-phonon interactions in graphene based on curved space hydrodynamics. Interactions of phonons  with charge carriers increase the electrical resistivity of the material. Our approach captures the  lattice vibrations as curvature changes in the  space through which electrons move following hydrodynamic equations. In this picture, inertial corrections to the electronic flow arise naturally effectively producing  electron-phonon interactions. The strength of the interaction is controlled by a coupling constant, which is temperature independent.  We apply this model to graphene and recover satisfactorily the linear scaling law for the resistivity that is expected at high temperatures. Our findings open up a new perspective of treating electron-phonon interactions in graphene, and also in other materials where electrons can be described by the Fermi liquid theory.
\end{abstract}


\maketitle

At finite temperatures,  phonons  interact with charge carriers, and therefore, contribute to the electrical resistivity of the respective material \cite{rossler2009solid}. 
In suspended graphene, the temperature dependence of the resistivity follows a linear increase due to electron-phonon interactions above 100 K. 
 \begin{figure} 
\includegraphics[trim=80 250 55 200,clip,width=\columnwidth] {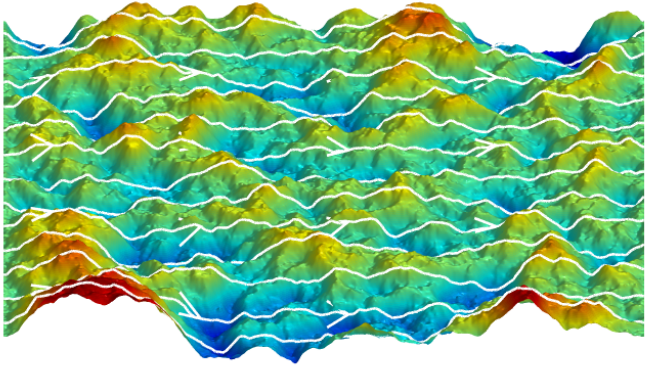}\caption{\label{fig:SheetWithStreamLines}
Graphene membrane at 300~K after the electronic flow has reached a steady state. Streamlines indicate the velocity field of the electronic flow that moves from left to right. The colours represent the height.}
\end{figure}
 \footnote{
 In graphene, the Debye temperature (corresponding to the theoretical highest phonon frequency in the material) is very high, $T \approx 2300$ K and $T \approx 1300$ K for planar and out-of-plane phonons, respectively \cite{PhysRevB.79.125416}. However, for low electron densities, the Fermi energy can be substantially smaller than the Debye energy, and only phonons with energy smaller than two times the Fermi energy, corresponding to a full backscattering of electrons, can scatter with the electrons. This  defines the Bloch-Gru\"neisen temperature $T_{BG}$ \cite{bloch1930elektrischen}, which for doped suspended graphene is around $100$ K \cite{Exp:ResistivityPropT,PhysRevB.77.115449}. 
The linear relation also holds when external strain is applied to the graphene samples \cite{PhysRevLett.105.266601}.}   
Recently it has been shown that  electronic flow in graphene can be modelled using relativistic hydrodynamic equations \cite{grapPRL,PoliniVisc,levitov2016electron,mendoza2013hydrodynamic,turbPRL,Oliver,PhysRevB.80.085109, HydroModelPhononsTraditional}.
Usually, in this hydrodynamic  formalism, electron-phonon interactions are included into the equation for the electrical current density $\vec{J}$ as a damping term, $-\vec{J}/\tau_p$, with a characteristic relaxation time $\tau_p$ \cite{PhysRevB.80.085109, HydroModelPhononsTraditional}. However, a proper inclusion of lattice-fluid interactions are missing. 
Here, we present a novel approach where we account for the electron-phonon interactions by including inertial corrections due to the deformations of the graphene sheet. 
We address the question whether these inertial corrections can   recover the linear temperature relation of the electrical resistivity in the high temperature regime.  

For this purpose, 
we use molecular dynamics simulations to simulate different graphene membranes at different temperatures and imposed strains. From the position of the atoms, we build the   coordinate system in which the electrons flow, and extract the inertial corrections to include them into the two-dimensional hydrodynamic equations. In Fig.\ref{fig:SheetWithStreamLines}, we can observe how the presence of thermal fluctuations produces  curved streamlines of the electron velocity field.

\begin{figure} 
\includegraphics[trim=450 120 255 200,clip,width=\columnwidth] {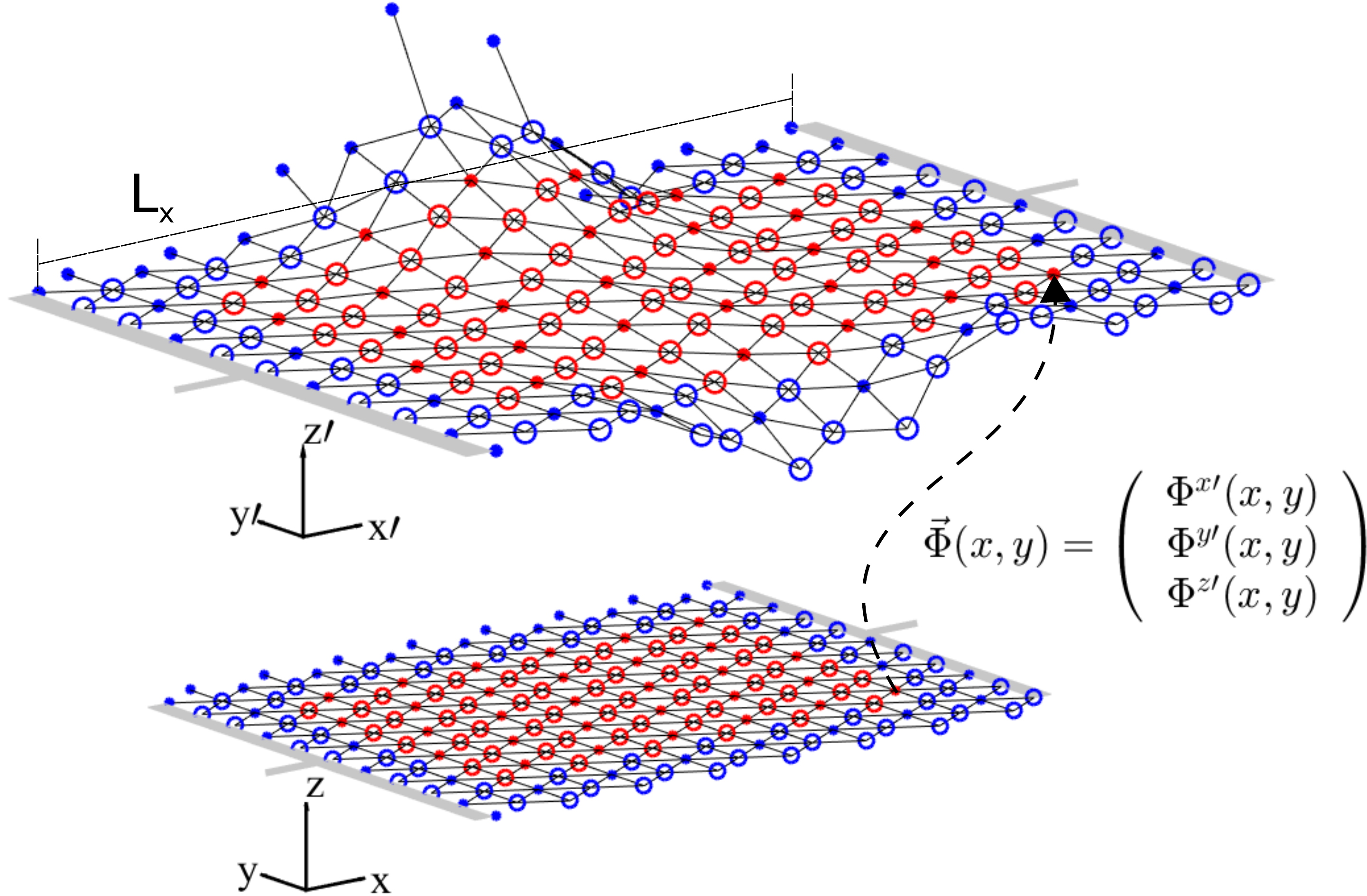}\caption{\label{fig:LBLattice} Discrete mapping (dashed arrow) that describes the embedding of the two-dimensional triangular lattice (bottom figure) in the three-dimensional Cartesian space (top figure). The circles represent the atoms.
The electrical contacts are shown in grey at both ends of the graphene sheet.}
\end{figure}

To simulate the graphene sheets with molecular dynamics we use the adaptive intermolecular reactive bond-order (AIREBO) potential \cite{stuart2000reactive}. This many-body potential has been developed to simulate molecules made of carbon and hydrogen atoms, and  
has been very successful to reproduce the phonon dispersion curves of graphene \cite{koukaras2015phonon}.  
In our simulations we consider suspended   graphene sheets  of $ L_x =99.2$~\AA, (see Fig.~\ref{fig:LBLattice}), but also two stretched cases  with $100.6$~\AA~and  $102.1$~\AA. As shown in Fig.~\ref{fig:LBLattice} the zigzag edge is located along the $x$-direction (top and bottom) and accordingly the armchair edge along the $y$-direction (left and right).  The atoms can freely move in all directions, except the ones at the left and right boundary, representing the electrical contacts (grey   in Fig.~\ref{fig:LBLattice}). The distance between contacts will be denoted by $L_x$. Note that since the left and right boundaries are fixed,   we are effectively applying strain to the graphene samples. We simulate several samples at different temperatures within the range of $100$-$600$ K controlling the temperature of the system with a   Nos\'e-Hoover thermostat. All molecular dynamics simulations run  (during the thermalisation procedure)   with a time step $\Delta t=1$~fs, which is small enough to capture the dynamics of the carbon-carbon interactions accurately. Randomised velocities are attributed to the atoms at the beginning of each simulation. After reaching thermal equilibrium we are ready to couple the graphene sheet to the fluid solver and introduce electrical currents. 

The electronic flow in flat graphene can be modelled using relativistic hydrodynamic equations, which are responsible for the conservation of particles, $\partial_\mu N^\mu=0$, and energy and momentum, $\partial_\mu T_0^{\mu \nu}=0$, where $N^\mu = n U^{\mu}$ and $T_0$ are the $3$-particle flow and the energy-momentum tensor, respectively. Here, $n$ is the number of particle density and $U^{\mu}$ the $3$-velocity of the electronic flow. The energy momentum tensor reads 
$T_0=(\epsilon +p) U^\mu U^\nu/v_F^2 -p \eta^{\mu \nu} +  \pi^{\mu \nu}$, 
where   $\eta_{\mu \nu}$ are the components of the Minkowski metric and $\pi^{\mu \nu}$ is the shear-stress tensor, which can be approximated by the equation $\pi^{\mu \nu} \approx \kappa (\eta^{\mu \lambda} \partial_\lambda U^\nu +\eta^{\nu \lambda}\partial_\lambda U^\mu)$, with $\kappa$ being the shear viscosity. $\epsilon$ and $p$ are   energy density and pressure, respectively. 
The equation of state completes this set of equations, which for the case of graphene is given by $\epsilon=2P$. Here, the Fermi velocity  is denoted by $v_F$. 
To take into account the curvature of suspended graphene sheets, we   consider inertial corrections in these equations. For this purpose, we first use the covariant formulation of the hydrodynamic equations in curved space to determine the terms responsible for the inertial corrections, and afterwards, we introduce them as a forcing term into the respective equations for flat graphene.  
The conservation equations in curved manifolds are derived by replacing the partial derivative by the covariant derivative and extending the energy-momentum tensor to curved space, i.e. $T^{\mu \nu}=T_0^{\mu \nu}+p (\eta^{\mu \nu}-g^{\mu \nu})$. 
The covariant derivative $\nabla_\mu$ in the curvilinear coordinate system can be expressed through   the respective partial derivative by $\nabla_\mu N^\mu=\partial_\mu N^\mu +\Gamma^\mu_{\mu \lambda} N^{\lambda }$, and $\nabla_\mu T^{\mu \nu}=\partial_\mu  T^{\mu \nu} +\Gamma^\mu_{\mu \lambda} T^{\lambda \nu}+ \Gamma^\nu_{\mu\lambda}  T^{\mu \lambda}$, with $\Gamma^\mu_{\mu \lambda}$ being the Christoffel symbols computed as follows: 
\begin{equation}
\label{eqn:ChristSymbols0}
 \Gamma^\sigma_{\mu \nu}=\frac{1}{2} g^{\sigma \rho}\left ( \frac{\partial g_{\rho \mu}}{\partial x^\nu}+\frac{\partial g_{\rho \nu}}{\partial x^\mu}-\frac{\partial g_{\mu \nu}}{\partial x^\rho} \right ) \quad .
\end{equation}

At this stage we include a temperature independent coupling constant $\alpha$ in front of the inertial corrections  which accounts for the strength of the electron-phonon interaction. Thus, the set of equations can be written as

 \begin{subequations} \label{eqn:ConsEqcnImplemented}
\begin{eqnarray}
   \partial_\mu N^\mu=\alpha F_N,
\end{eqnarray}
\begin{eqnarray}
\partial_\mu  T_0^{\mu \nu}=\alpha  F_T^\nu, 
\end{eqnarray}  
\end{subequations}
 with 
\begin{subequations} \label{eqn:ForcingTerms}
\label{eqn:ForcesFromMetric}
\begin{eqnarray}
  F_N=-\Gamma^\mu_{\mu \lambda} N^{\lambda }
\end{eqnarray}
\begin{eqnarray}
F_T^\nu= -\Gamma^\mu_{\mu \lambda} T_0^{\lambda \nu}- \Gamma^\nu_{\mu\lambda}  T_0^{\mu \lambda}
-(\eta^{\mu\nu}-g^{\mu \nu})\partial_\mu p. 
\end{eqnarray}  
\end{subequations}

To solve the hydrodynamic equations, we use the lattice Boltzmann solver described in Ref.~\cite{Oettinger, Oliver}, and add the inertial contributions as external forces (see Supplementary Information). 
The model discretises   space   as regular triangular lattice, which we couple to graphene's hexagonal lattice.
 As shown in Fig. \ref{fig:LBLattice}, the atoms form a two-dimensional manifold which can be described by a discrete mapping $\Phi(ct,x,y)$ from the  curved space to the three-dimensional flat space (reference frame of the laboratory, where the metric is given by the Minkowski-metric).  
 The metric tensor can be computed by
\begin{equation}
g_{\mu \nu}=\frac{\partial \Phi^\alpha ( ct,x,y )}{\partial x^\mu}\frac{\partial \Phi^\beta(  ct,x,y )}{\partial x^\nu} \eta_{\alpha \beta}.
\end{equation}
Similarly, the Christoffel symbols are obtained from Eq.~\eqref{eqn:ChristSymbols0}. Further details on this calculations can be found in the Supplementary Information. The graphene sheet possesses zigzag boundaries at the left and right end.
For the fluid solver we impose periodic boundary conditions at these boundaries, which   correspond  to the in- and outlet. The free sides of the graphene sheet possesses armchair geometry and, therefore, we impose free slip boundary conditions.
 
In order to obtain general results and allow for more insight, we make the hydrodynamic equations   dimensionless  (See Supplementary Information for the detailed analysis). Considering that the shear-stress tensor depends linearly on the shear viscosity  \cite[p. 109]{Book:RelaBoltEqua} we find that the physics of the electronic flow only depends on two dimensionless numbers: $A_1=\kappa_0 v_F'^2 u_0/x_0 \epsilon_0$, and
$A_2=e_0 E_0/x_0 \epsilon_0 v_F'^2$, where $\epsilon_0$, $u_0$, $x_0$, $\kappa_0$, $e_0$, and $E_0$ are characteristic values for the energy density, velocity, length, shear viscosity, electric charge, and electric field, respectively. Note that $1/A_1$ is related to the Reynolds number \cite{pozrikidis2009fluid}. The parameter range of our  simulations is $0.14 \leq A_1 \leq 2.05$ and  $10^{-8} \leq A_2 \leq 10^{-6}$.
 
To produce an electrical current, we apply an external electric field in $x$ direction, $E^x=1.43 \times 10^9 - 10^{11}$~V/m. 
In our simulations, we  set   $n=1.36\times 10^{13}$~/cm$^2$, chemical potential $\mu=\hbar v_F\sqrt{\pi n}\approx 6.89\times 10^{-20}$~kgm$^2$/s$^2\approx 0.43$~eV, and $p=\frac{1}{3} \mu n \approx 3.12 \times 10^{-3}$~kg/s$^2$. 
We use the shear viscosity  for doped graphene given by \cite{PoliniVisc}
\begin{equation}
\kappa(T) = c_\nu n (\frac{\mu}{k_BT})^\frac{3}{2},
\end{equation}
with $c_\nu \approx 1.33 \times 10^{-34} \text{kgm}^2\text{/s} \approx \frac{5}{4}\hbar$.
We couple the electronic fluid to the atomistic simulation  using the same length-scale $x_0=\Delta x_{MD}=1$~\AA~ and time scale $t_0=\Delta t_{MD}=10^{-5}$~fs.
After each iteration we compute the metric tensor and the Christoffel symbols and simulate the electronic flow until we  obtain the electrical current in steady state. 
The momentum change in the fluid is imposed on the atoms to ensure  momentum   conservation of the system.

\begin{figure} 
\includegraphics[trim=35 145 80 160, clip,width=\columnwidth] {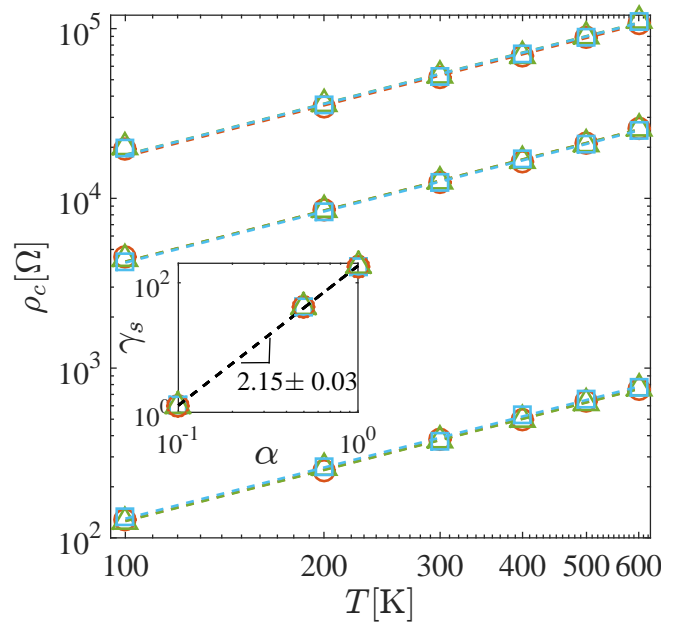}\caption{\label{fig:ResistivityLogDifferentAlpha}
Temperature dependence of the electrical resistivity of graphene due to inertial corrections for different coupling constants $\alpha$. 
 The dashed lines correspond to the best fit for the function $\rho = \gamma_s T + \rho_0=\rho_c+\rho_0$ for $\alpha=\lbrace 1.0, 0.5,  0.1\rbrace$ with $\gamma_s(\alpha=1.0)=180 \pm 20~\Omega$/K,$\gamma_s(\alpha=0.5)=44 \pm 7~\Omega$/K and $\gamma_s(\alpha=0.1)=1.3 \pm 0.1~\Omega$/K. The  colors represent graphene membranes with different strains where red circles stand for a graphene sheet with size $L_x=99.2$ \AA, green triangles for $L_x=100.6$ \AA~and blue squares for $L_x=102.1$ \AA. Inset: Log-log plot of $\gamma_s$ dependence on $\alpha$. In both figures, the errorbars are smaller than the  symbols. }
\end{figure}

We perform simulations for different strains and different values of the coupling constant and measure the resistivity of the graphene sheets using Ohm's law,
$\langle \rho \rangle = E^x L_y/I(t \to \infty)$, where the electrical current is given by 
 \begin{equation}
 \label{eqn:Current}
I(t)= \int_0^{L_y} e n(ct,x=L_{x},y)  U^x(ct,x=L_x,y) dy.
\end{equation} 
In Fig.~\ref{fig:ResistivityLogDifferentAlpha}, we observe that all our calculations of the electrical resistivity of graphene exhibit a linear dependence with temperature, $\rho = \gamma_s(\alpha) T+\rho_0(\alpha, L_x)$. This linear dependence is in agreement with experimental measurements and theoretical predictions \cite{Exp:ResistivityPropT, PhysRevB.77.115449}. Additionally, the slope $\gamma_s(\alpha)$ only depends on the strength of the coupling constant $\alpha$ but not on the applied strain, which is also in agreement with previous works \cite{PhysRevLett.105.266601}. We can also observe from the inset of Fig.~\ref{fig:ResistivityLogDifferentAlpha} that $\gamma_s\propto  \alpha^{2.15 \pm 0.03}$. 
The factor $\gamma_s$ can be compared with the values from the Boltzmann transport theory \cite{PhysRevB.77.115449}, where  
\begin{equation}
\label{eqn:ResistivityAnalyticalExpression}
\rho = \gamma T \equiv \frac{\pi D^2 k_B}{4e^2\hbar \rho_m v^2_{ph}v_F^2 } T,
\end{equation}
with $\rho_m=7.6\times 10^{-7}$ kg/m$^2$ being the mass density of graphene, $v_{ph}$ the phonon velocity and $D$ the deformation-potential coupling constant. Experimental and theoretical results suggest that the deformation-potential varies within the range of $4.5 \leq D \leq 50$ eV \cite{borysenko2010first,chen2008intrinsic,bolotin2008temperature},
and phonon velocities within $2\times 10^4 \leq v_{ph} \leq 3 \times 10^4$ m/s. Thus, one expects $\gamma$ to be in the range $0.012 \leq \gamma \leq 2.8~\Omega$/K. 
From our results we deduce that   $0.01\leq \alpha \leq 0.14$ to obtain results compatible with $\gamma$. 
 The residual resistivity $\rho_0$ is $704\pm 33~\Omega$ for $\alpha=0.1$ and    is a consequence of the static corrugations (ripples)  present at any temperature. These ripples stabilize the two-dimensional crystal circumventing the Mermin    Wagner theorem, which states that crystalline order cannot exist in two dimensions \cite{mermin-wagner}.

\begin{figure}  
\includegraphics[trim=20 110 55 150,clip,width=\columnwidth] {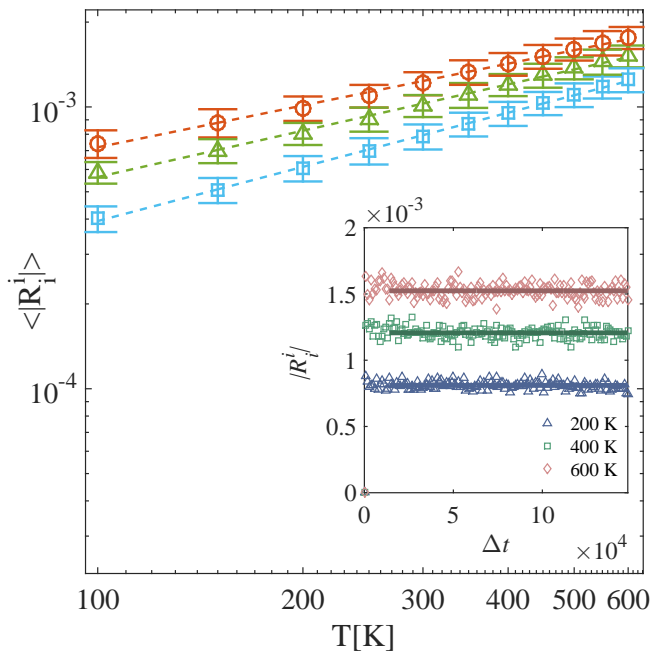}\caption{\label{fig:TimeAveragedRicciScalarvsTmp}
Maximal absolute value of the Ricci scalar $|R^i_i|$  for graphene membranes with distances $L_x$. Main panel: time averaged Ricci scalar $\langle |R^i_i| \rangle$ for different temperatures and   distances $L_x$. Red circles correspond to $L_x=  99.2$ \AA, with an exponent $0.49\pm 0.02$.  Green triangles  correspond to $L_x=  100.6$ \AA, with an exponent $0.55\pm 0.03$. 
Blue squares  correspond to $L_x=  102.1$ \AA, with an exponent $0.64\pm 0.01$. 
Inset panel: Time evolution of $|R^i_i|$  for graphene membranes at different temperatures for   $L_x=  102.1$ \AA.}
\end{figure}
The existence of a resistivity points to the presence of dissipation.
Recently, it has been shown that energy dissipation in flows through curved manifolds arises from the curvature \cite{JD:CurvatureInduceDissip}.
 The Ricci scalar (or curvature scalar) is a measure for the curvature and is given by
$R^\mu_\mu=g^{\nu \sigma}(\Gamma^\mu_{\nu\sigma,\mu}-\Gamma^\mu_{\nu\mu,\sigma}+\Gamma^\delta_{\nu\sigma}\Gamma^\mu_{\mu\delta} +\Gamma^\delta_{\nu\mu}\Gamma^\mu_{\sigma\delta})$. 
We have analysed the Ricci scalar for different strains and temperatures and found that its maximal absolute value remains constant during   time evolution and depends only on the temperature of the system (see inset of Fig.~\ref{fig:TimeAveragedRicciScalarvsTmp}). 
As   function of temperature the Ricci scalar follows a  power-law with an exponent  that increases   with the applied strain (see main panel in Fig.~\ref{fig:TimeAveragedRicciScalarvsTmp}).

We have also studied the standard deviation   of the heights  of our graphene sheets, and found that they are virtually  not   influenced by temperature, within the range of $100-600$ K (see Fig.~\ref{fig:HeightSTD} of Supplementary Information). Therefore, we conclude that,   an increase of temperature mainly induces in-plane atomistic motion, increasing the   average effective bond length (see Figs.~\ref{fig:DistrBL} a-c in the Supplementary Information), and consequently, increasing the local curvature. Since curvature induces    energy dissipation \cite{JD:CurvatureInduceDissip,JD:PoiseuilleFlow}, this explains why larger temperatures result in larger electrical resistivities.

To summarise, we have shown that inertial corrections lead to a linear dependence of the electrical resistivity with temperature in suspended graphene, which is in agreement with experimental measurements of the electrical resistivity due to electron-phonon interactions at high temperatures. To characterize the strength of the electron-phonon interaction in our model, we have introduced a coupling constant $\alpha$, and for values of $\alpha \approx 0.1$ one recovers the same values of $\gamma$ known from theoretical predictions and experimental measurements. Additionally, the resistivity does not depend  on the strain imposed to the sample, which is also in agreement with previous work \cite{PhysRevLett.105.266601}. We also analysed the maximal absolute value of the curvature as function of temperature and found a power-law behaviour with an exponent   that changes slightly with the applied strain. Finally, by studying the average height fluctuations of the graphene sheets, we also discovered that by increasing the temperature,  the height fluctuations remain almost unchanged, and consequently, including larger local values of curvature. In our model, larger curvature implies more energy dissipation, and consequently, larger electrical resistivity.
 
Our finding opens up many interesting questions, as for instance, if one can  discern the influence of different types of vibrational phonons, e.g. flexural, acoustic, and optical, by studying how they introduce curvature into the hydrodynamic system. By extending our molecular dynamics simulations, one could also explore the electrical resistivity due to the electron-phonon interactions when the graphene sample is placed on a substrate \cite{chen2008intrinsic}. 

Finally, our approach can also be applied to other two- and three-dimensional crystals, where electrons are well described as Fermi liquid. In those structures, ion displacements due to phonons can induce intrinsic curvature, and consequently, energy dissipation in the electronic flow. Applying our model to other materials will be an interesting subject for future research.

\begin{acknowledgments}
We thank   the European Research Council (ERC) Advanced Grant No. 319968-FlowCCS for financial support.
\end{acknowledgments}


%

\appendix

\section{Conservation equations for curved spaces}
The lattice Boltzmann model for flat space reported in Ref.~\cite{Oettinger} needs to be extended in order to simulate relativistic hydrodynamics in curved spaces. This numerical model solves the following conservation equations:
\begin{subequations} \label{eqn:ConservationFlat}
\begin{eqnarray}
  \partial_\mu N^\mu=0,
\end{eqnarray}
\begin{eqnarray}
\partial_\mu T_0^{\mu \nu}=0,
\end{eqnarray}  
\end{subequations}
where $T_0^{\mu\nu}$ is the energy momentum tensor for flat space,
\begin{equation} 
T_0^{\mu \nu}= (\epsilon +p)\frac{U^\mu U^\nu}{v_F^2} -p \eta^{\mu \nu} + \pi^{\mu \nu} .
\end{equation}

In curved spaces, the energy momentum tensor has to be conserved as well. However, conservation of quantities in a curved manifold needs to be expressed through covariant derivatives, instead of partial derivatives. The conservation equations for the number of particles and energy momentum tensor are given by
\begin{subequations} \label{eqn:ConservationCurved}
\begin{eqnarray}
  \nabla_\mu N^\mu=0,
\end{eqnarray}
\begin{eqnarray}
\nabla_\mu T^{\mu \nu}=0,
\end{eqnarray}  
\end{subequations} 
where the covariant derivatives $\nabla_\mu$ of the 3-particle flow and energy-momentum tensor relate to the partial derivatives $\partial_\mu$ as follows: 
\begin{subequations} \label{eqn:CovariantDerivTrafos}
\begin{eqnarray}
  \nabla_\mu N^\mu=\partial_\mu N^\mu +\Gamma^\mu_{\mu \lambda} N^{\lambda },
\end{eqnarray}
\begin{eqnarray}
\nabla_\mu T^{\mu \nu}=\partial_\mu  T^{\mu \nu} +\Gamma^\mu_{\mu \lambda} T^{\lambda \nu}+ \Gamma^\nu_{\mu\lambda}  T^{\mu \lambda}, 
\end{eqnarray}  
\end{subequations}
 where $\Gamma^\mu_{\mu \lambda}$ denote the Christoffel symbols which are computed with the metric tensor using 
\begin{equation}
\label{eqn:ChristSymbols}
 \Gamma^\sigma_{\mu \nu}=\frac{1}{2} g^{\sigma \rho}\left ( \frac{\partial g_{\rho \mu}}{\partial x^\nu}+\frac{\partial g_{\rho \nu}}{\partial x^\mu}-\frac{\partial g_{\mu \nu}}{\partial x^\rho} \right ) \quad .
\end{equation}
Additionally, the energy momentum tensor in curved manifolds reads 
\begin{equation} 
T^{\mu \nu}= (\epsilon +p)\frac{U^\mu U^\nu}{v_F^2} -p g^{\mu \nu} +  \pi^{\mu \nu},
\end{equation}
where $g^{\mu \nu}$ is the metric tensor and $\pi^{\mu \nu}$ the shear-stress tensor, which can be approximated by the equation $\pi^{\mu \nu} \approx \kappa (g^{\mu \lambda} \nabla_\lambda U^\nu +g^{\nu \lambda}\nabla_\lambda U^\mu)$. Using the property that $\nabla_\mu g^{\mu\nu}=0$, we can rewrite the covariant derivative of the energy-momentum tensor as 
\begin{equation}
\label{eqn:nablaTc2nablaT}
\nabla_\mu T^{\mu \nu}=\nabla_\mu T_0^{\mu \nu} +(\eta^{\mu\nu}-g^{\mu \nu})\nabla_\mu p .
\end{equation}
Thus the conservation equations \eqref{eqn:ConservationCurved} expressed with the energy-momentum tensor for the flat space $T_0^{\mu \nu}$ reads  
\begin{equation}
\label{eqn:ConservationTc_expressedwith_T}
\begin{aligned}
 0&=\nabla_\mu T^{\mu \nu}=\nabla_\mu T_0^{\mu \nu} +(\eta^{\mu\nu}-g^{\mu \nu})\nabla_\mu p\\
 &=\partial_\mu  T_0^{\mu \nu}+\Gamma^\mu_{\mu \lambda} T_0^{\lambda \nu}+ \Gamma^\nu_{\mu\lambda}  T_0^{\mu \lambda}
+(\eta^{\mu\nu}-g^{\mu \nu})\partial_\mu p,
\end{aligned}
\end{equation}
where we have used the fact that partial derivatives equal covariant derivatives when acting on scalars. At this stage we also include a temperature independent coupling constant $\alpha$ which accounts for the strength of the inertial corrections. Thus, this set of equations can be written as

 \begin{subequations} \label{eqn:ConsEqcnImplemented}
\begin{eqnarray}
   \partial_\mu N^\mu=\alpha F_N,
\end{eqnarray}
\begin{eqnarray}
\partial_\mu  T_0^{\mu \nu}=\alpha  F_T^\nu, 
\end{eqnarray}  
\end{subequations}
 with 
\begin{subequations} \label{eqn:ForcingTerms}
\label{eqn:ForcesFromMetric}
\begin{eqnarray}
  F_N=-\Gamma^\mu_{\mu \lambda} N^{\lambda }
\end{eqnarray}
\begin{eqnarray}
F_T^\nu= -\Gamma^\mu_{\mu \lambda} T_0^{\lambda \nu}- \Gamma^\nu_{\mu\lambda}  T_0^{\mu \lambda}
-(\eta^{\mu\nu}-g^{\mu \nu})\partial_\mu p. 
\end{eqnarray}  
\end{subequations}
Here $F_T^\nu$ and $F_N$ can be introduced in the numerical model as external forces using the technique described in Ref.~\cite{Oliver}.

\section{Dimensionless quantities}
\label{Sec:DimensionlessEqnDerivation}
In order to obtain general results we transform the hydrodynamic equations to dimensionless form and determine the dimensionless quantities that characterise our systems. We also include a force density $\vec{F}^{ext}=ne\vec{E}$ describing an external electric field $\vec{E}$ to the energy-momentum conservation equation \eqref{eqn:ConsEqcnImplemented}.  

To write the equations in a dimensionless form, we first express all   relevant quantities in dimensionless form: $x=x_0 x', u=u_0u',t=t_0t',\epsilon=\epsilon_0 \epsilon',E=E_0E'$,  $\kappa=\kappa_0 \kappa'$ $e=e'e_0$,
 $n=n'/x_0^2$ and 
$v_F=u_0 v_F'$ where all primed variables are dimensionless. Furthermore, the relation $u_0t_0=x_0$ holds, which allows us to write the temporal derivative $\partial_0=(1/v_F)\partial_t= (1/v_F'u_0t_0)\partial_{t'}=(1/v_F'x_0)\partial_0'$, and thus for $v_F'=1$, the relation $\partial_\mu=(1/x_0)\partial_\mu'$ is also satisfied.

We divide the energy-momentum tensor in an equilibrium part and a dissipative part $T^{\mu\nu}=T^{\mu \nu (eq)} +\pi^{\mu \nu}$. The equilibrium part can be written in its dimensionless form $ T^{\mu \nu (eq)}=\epsilon_0 T'^{\mu \nu (eq)} $. Assuming that the shear-stress tensor is linearly dependent on the shear viscosity $\kappa$ \cite[p. 109]{Book:RelaBoltEqua} and taking into account that the shear stress tensor has the same units as the energy-momentum tensor ($\kappa_0=\frac{\epsilon_0x_0}{u_0}$), we get $ \pi^{\mu \nu}=\kappa_0 (\epsilon_0/\kappa_0) \pi'^{\mu \nu}=\kappa_0 (u_0/x_0) \pi'^{\mu \nu}$. The force densities \eqref{eqn:ForcesFromMetric}  transform as  $F_N = (\epsilon_0/x_0) F_N'$ and $F_T^\nu = (\epsilon_0/x_0) F_T'^{\nu}$, respectively. This can be derived from the fact that, in our case, the metric is dimensionless and thus the Christoffel symbols (as derivative of the metric) transform as $\Gamma^{\mu \nu}_\sigma=(1/x_0) \Gamma'^{\mu \nu}_\sigma$.  

With the considerations above, equations \eqref{eqn:ConsEqcnImplemented} read

\begin{subequations} \label{eqn:ConsEqDimensionlessDeriv}
\begin{eqnarray} 
   \frac{1}{x_0^3} \partial_\mu' N'^{\mu} = \frac{1}{x_0^3} F'_N,
\end{eqnarray}
\begin{eqnarray}
\frac{\epsilon_0}{x_0} \partial_\mu' T'^{\mu \nu (eq)} &+\frac{\kappa_0 u_0}{x_0^2} \partial_\mu'  \pi'^{\mu \nu (eq)}\\ &=\frac{\epsilon_0}{x_0}  F_T'^{\nu}+\frac{e_0 E_0}{x_0^2} n'e'E'^{\nu}. 
\end{eqnarray}  
\end{subequations}
While the first equation remains unchanged, we observe after multiplying by $(x_0/\epsilon_0)$ that the second equation only depends on two dimensionless numbers 
\begin{eqnarray}
A_1=\frac{\kappa_0 u_0}{x_0 \epsilon_0 } \quad ,
\end{eqnarray}
and
\begin{eqnarray}
A_2=\frac{e_0 E_0}{x_0 \epsilon_0} \quad .
\end{eqnarray}  

\section{Strain}

\begin{figure}
  \includegraphics[trim=50   140  50 140, clip,width=\columnwidth]{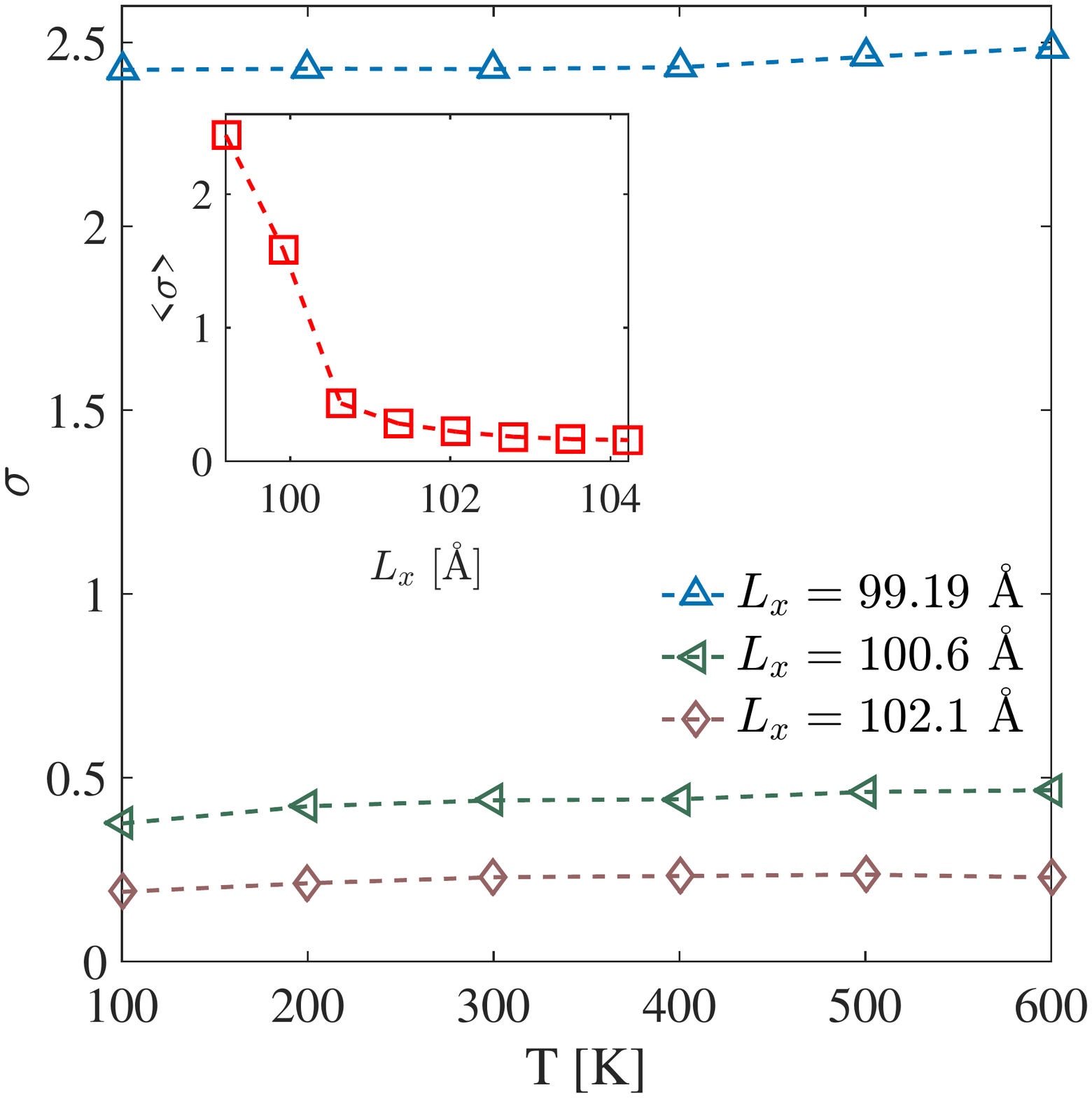}  \caption{Main figure: Standard deviation of the heights $\sigma$ as a function of temperature for different  $L_x=\lbrace 99.2, 100.6, 102.1\rbrace$.
Inset: Temperature averaged standard deviation $\langle \sigma \rangle$ for different distances $L_x$.    } 
\label{fig:HeightSTD}
\end{figure}  
 
All simulations are performed with the same number of carbon atoms ($N=4114$).   The atoms at the left and right boundary are fixed at a distance $L_x$ from each other and are not allowed to move nor change their effective bond lengths. The simulations are preformed for different       $L_x=99.2-104.2$~\AA~  which correspond to   initial effective bond lengths $a= 1.38-1.45$~\AA.   
 
%
%
%
\begin{figure}
\begin{center}
\subfigure[]{
    \includegraphics[trim=50  180  140 200, clip,scale=0.47]{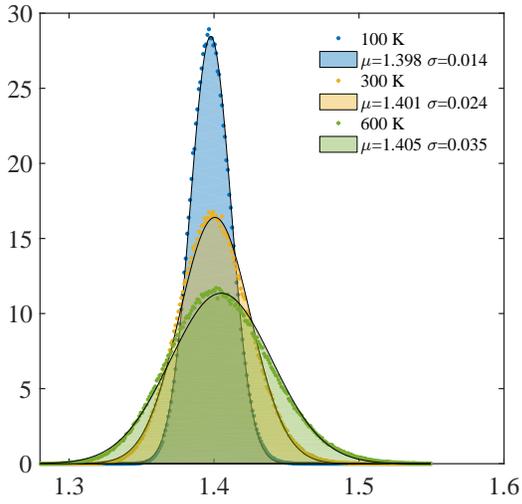}    \label{fig:DistrBL138} 
  }
  \subfigure[]{
\includegraphics[trim=50  180  140 200, clip,scale=0.47]{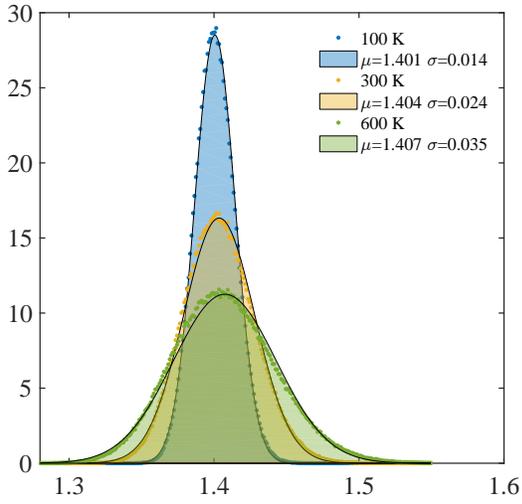} \label{fig:DistrBL140} 
}
 \subfigure[]{
\includegraphics[trim=50  180  140 200, clip,scale=0.47]{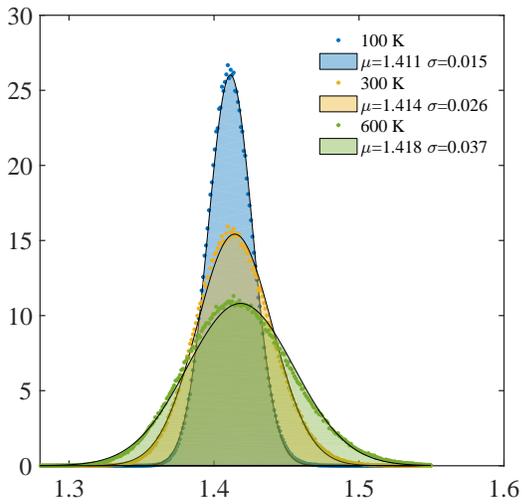} \label{fig:DistrBL142} 
}
\caption{\label{fig:DistrBL} 
Distribution of the effective bond lengths for     $L_x=99.2$ \AA~\subref{fig:DistrBL138}, 
 $L_x=100.6$ \AA~\subref{fig:DistrBL140} and
 $L_x=102.1$ \AA~\subref{fig:DistrBL142}. The black line corresponds to a Gauss distribution with mean $\mu$ and standard deviation $\sigma$.
}
\end{center}
\end{figure}

The distance $L_x$ is related to the tension imposed on the graphene membrane.   
One expects that increasing $L_x$ results in a more stretched membrane and the height fluctuations are reduced.  
Figure \ref{fig:HeightSTD} confirms that the height fluctuations $  \sqrt{\langle z^2\rangle -\langle z\rangle^2}$ depend on $L_x$ and therefore also on the strain imposed on the graphene membrane. Interestingly, the height fluctuations are  independent on the temperature. 
In contrast, the width of the effective bond length distribution presents a strong temperature dependence  as shown in Figs. \ref{fig:DistrBL} a-c. 

This finding is related to the Ricci scalar analysis (see Fig. \ref{fig:TimeAveragedRicciScalarvsTmp}), where it has been shown that the absolute value of the Ricci scalar has a much stronger temperature dependence than a dependence on $L_x$. A higher temperature results in  a broader width of the effective bond length distribution. This leads to a higher average Ricci scalar and consequently to more curvature in the system. The curvature introduces shear which is responsible for saturation of the current.

\end{document}